\documentclass[pra, 12 pt]{revtex4}
\usepackage[english]{babel}
\usepackage{amsmath}
\usepackage{epsfig}

\begin{document}


\title{METAL - NON-METAL TRANSITION AND THE SECOND CRITICAL POINT IN EXPANDED METALS}

\author{V.B. Bobrov $^{1,2}$, S.A. Trigger $^{1,3}$, A.G. Zagorodny $^{4}$}

\address{$^1$Joint\, Institute\, for\, High\,
Temperatures, Russian\, Academy\, of\, Sciences, 13/19, Izhorskaia
Str., Moscow\, 127412, Russia;\\
$^2$National Research University "MPEI"\,,
Krasnokazarmennaya str. 14, Moscow, 111250, Russia;\\
$^3$Eindhoven\, University\, of\, Technology, P.O.\, Box\, 513, MB\,
5600\, Eindhoven, The\, Netherlands;\\
$^4$ Bogolyubov Institute  for Theoretical Physics,
National Academy of Sciences of Ukraine,
14B Metrolohichna Str., Kiev 03143, Ukraine.\\
emails:\, vic5907@mail.ru,\;satron@mail.ru, Zagorodny@nas.gov.ua}

\begin{abstract}
Based on the non-relativistic Coulomb model within which the matter is a system of interacting electrons and nuclei, using the quantum field theory and linear response theory methods, opportunity for the existence of the second critical point in expanded metals, which is directly related to the metal--nonmetal transition, predicted by Landau and Zeldovitch, is theoretically justified. It is shown that the matter at the second critical point is in the state of true dielectric with zero static conductivity. The results obtained are in agreement with recent experiments for expanded metals. The existence of the second critical point is caused by the initial multi-component nature of the matter consisting of electrons and nuclei and the long-range character of the Coulomb interaction.\\

PACS number(s): 64.60.F-, 64.60.fd, 67.10.Fj, 65.40.Ba\\
\end{abstract}.

\maketitle

Currently, the experimental data on the transition of liquid metals at high temperatures and pressures to the non-metal state were obtained not only for mercury, cesium, and rubidium (see [1] for more details), but also, due to dynamic experiments, for aluminum, iron, nickel, copper, molybdenum, and tungsten (see [2] and references therein). However, the experimental study of this phenomenon called the metal-nonmetal (M-NM) transition for a long time did not confirm the old idea by Landau and Zeldovitch [3] about that the M-NM transition is the first-order phase transition (see Fig. 1). In this case, as is known [4], the M-NM phase transition, being the first-order transition, is characterized by:\\
(i) different densities (in metal and non-metal phases);\\
(ii) transition energy (heat);\\
(iii) critical point which does not coincide with the ordinary critical point of the liquid-vapor (L-V) phase transition;\\
(iv) triple point: liquid-liquid-vapor or vapor-vapor-liquid.

The absence of direct experimental data confirming the Landau and Zeldovitch [3] hypothesis from the viewpoint of testing the above items (i)-(iv) is caused by extraordinary difficult arrangement of static experiments at high temperatures and pressure. Therefore, strictly speaking, a final conclusion about the existence of this phase transition has not yet been made. As before, it seems most promising to study mercury to confirm the existence of the M-NM phase transition, for which this transition, judging by experimental data on the static conductivity, occurs at the lowest temperature and pressure among other liquid metals. Exactly for expanded liquid mercury in the region of the M-NM transition, small angle X-ray scattering measurements [6] and very accurate ultrasound velocity measurements [7] were performed. These measurements indirectly confirm the hypothesis of the correspondence of the M-NM transition to the first-order phase transition (for more details, see [5] and references therein). In this respect, no less significant are the results of [8] in which a technique for determining the equation of state was developed in studying the M-NM transition in expanded liquid iron by the dynamic method. This allowed the authors of [8] to draw the conclusion on the confirmation of the Landau and Zeldovitch hypothesis and the existence of the second critical point whose thermodynamic parameters differ from those of the ordinary critical point.
\begin{figure}
\includegraphics[width=8cm]{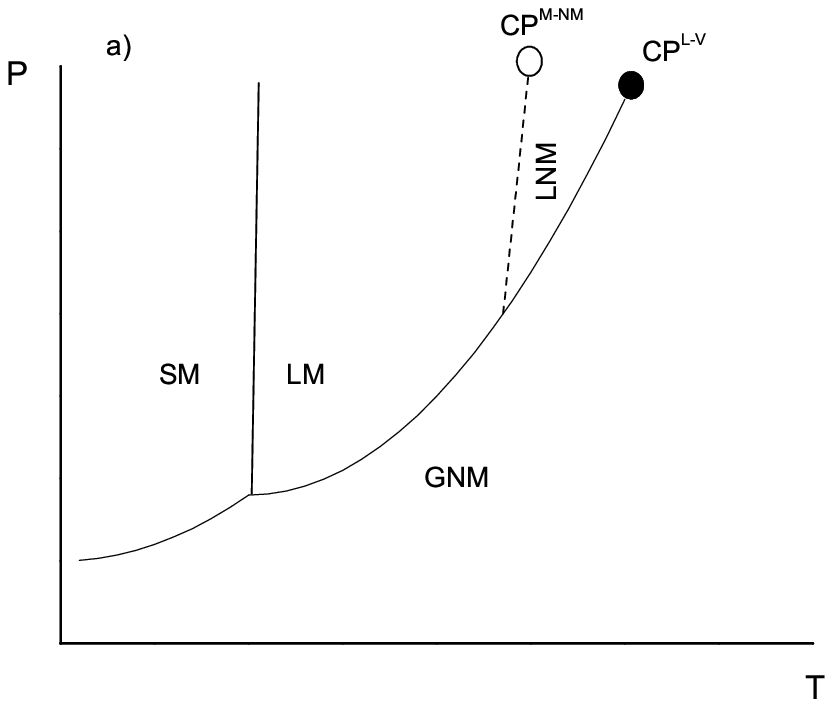}
\includegraphics[width=8cm]{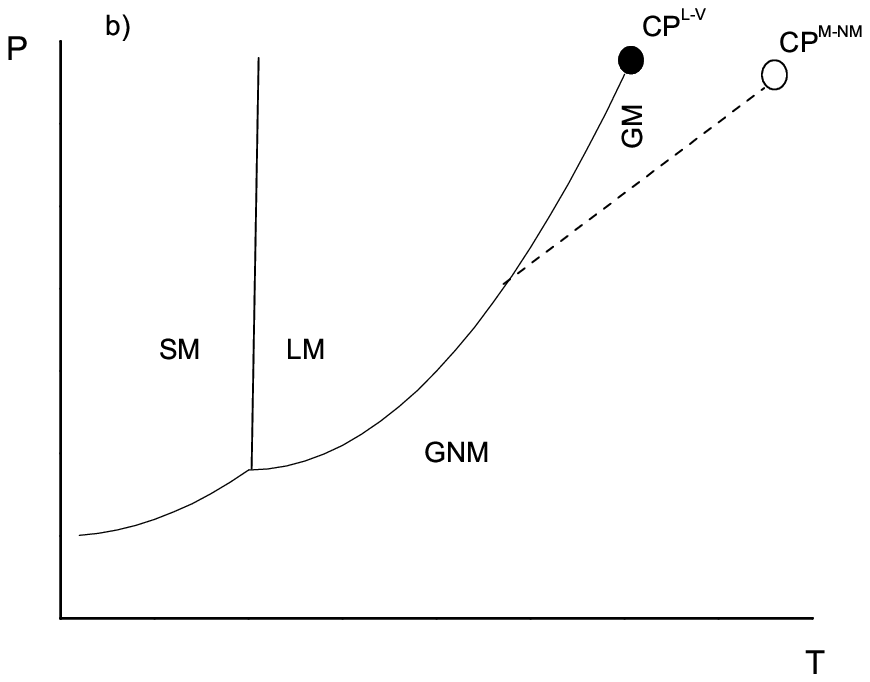}
\caption{Possible phase diagrams a) and b) with two critical points - for the liquid-vapor phase transition CP$^{L-V}$ and for the metal-non-metal phase transition CP$^{M-NM}$, according to Landau-Zeldovitch hypothesis [3]; GNM - non-metallic gas, GM - metallic gas, LNM- non-metallic liquid, LM - metallic liquid, SM - metallic solid.}
\label{fig.1}
\end{figure}

For this reason, it becomes extremely significant to construct the theory allowing a consistent explanation of the existence of the second critical point and its correspondence to the M-NM transition. It is clear that the most adequate basis of such a theory is the so-called Coulomb (or physical) model of the matter. This model considers a quantum non-relativistic system of electrons and nuclei interacting with each other by pure Coulomb interaction (Coulomb system, CS). For such a system the quasi-neutrality condition
\begin{eqnarray}
\sum_a z_a e n_a=0.
\label{B1}
\end{eqnarray}
is satisfied. Here $n_a$ is the average density of the number of particles of type $a$, which are characterized by the charge $z_{a}e$, mass $m_a$, and chemical potential $\mu_a$.

The main feature of the CS is the long-range nature of the interaction between charged particles, which simultaneously is the main advantage of the Coulomb model of matter. Such an approach does not require the use of model potentials (pseudopotentials) with fitting parameters, since the Coulomb interaction potential between matter particles is completely defined.

To construct the statistical theory of the CS, there exists a well developed formalism based on the quantum field theory methods; however, specific results for thermodynamic functions and kinetic coefficients mostly can be obtained only for the case of weakly nonideal systems (see, e.g., [9] and references therein). Nevertheless, there exist a number of exact results [10-13] for solving the problem of the second critical point and its interrelation with the static conductivity. Using these results, we will show that two critical points can exist in the two-component CS consisting of electrons (subscript e) and nuclei of the one type (subscript c). At one of the critical points, the matter is in the specific new state of "true dielectric" (state with zero static conductivity).

To prove this statement, we will proceed from the determination of the thermodynamic parameters of the critical point [4]
\begin{eqnarray}
K_T= - \frac{1}{V}\left(\frac{\partial P}{\partial V}\right) _{\{ N_a\},T} = \left\{\sum_{a,b=e,c} n_a n_b \left( \frac{\partial \mu_a}{\partial n_b} \right)_{n_d,T}\right\}^{-1} \rightarrow \infty.
\label{B2}
\end{eqnarray}
where $K_T$ is the isothermal compressibility of the considered two-component Coulomb system in the volume $V$ at temperature $T$, and pressure $P$, $N_a$ is the total average number of particles of type $a$. Relation (2) should be understood in the thermodynamic limit: $V\to \infty, N_a\to \infty, n_a = N_a/V=const$ while satisfying the quasi-neutrality condition (1).

We emphasize that, despite the quasi-neutrality condition, both densities $\{n_a \}$ and chemical potentials $\{\mu_a \}$ in the CS should be considered as independent parameters until deriving the final expression for the physical quantity characterizing the CS. Only in this final expression, the quasi-neutrality condition (1) is used (see [13,14] for more details). We note that from the stability requirement the Coulomb system can be considered only within quantum statistics [9]. In this case, from the dimension reasons, the CS critical parameters are purely quantum quantities [10]. For this reason, in the theoretical study, it is necessary to use specified chemical potentials $\{\mu _a \}$ instead of specified densities $\{n_a \}$ when defining the pressure $P$, i.e., the grand canonical distribution [13,14] should be used instead of the canonical distribution. This means that we should use derivatives $\left(\partial n_a/\partial \mu_b\right)_{\mu _d,T}$ instead of derivatives $\left(\partial\mu_a/\partial n_b \right)_{n_d,T}$. To this end, we use the thermodynamic equality
\begin{eqnarray}
\sum_{p}\left(\frac{\partial\mu_a}{\partial n_p} \right)_{n_d,T} \left(\frac{\partial n_p}{\partial \mu_b}\right)_{\mu _f,T}= \delta_{a,b}
\label{B3}
\end{eqnarray}
For the two-component CS, using the set of equations (3), it is easy establish the relation between derivatives $\left(\partial \mu_a/\partial n_b \right)_{n_d, T}$ and $\left(\partial n_a/\partial \mu_b \right)_{\mu _d,T}$. In this case, the isothermal compressibility takes the form [11-13]
\begin{eqnarray}
K_T=\left[  \left(\frac{\partial n_e }{\partial \mu _e } \right)_{\mu _c,T}
\left(\frac{\partial n_c }{\partial \mu _c} \right) _{\mu _e,T} -
\left(\frac{\partial n_e }{\partial \mu _c } \right)_{\mu _e,T}
\left(\frac{\partial n_c }{\partial \mu _e } \right) _{\mu _c,T}\right] \times
\left\{\varphi(\mu _e, \mu _c,T)   \right\}^{-1},
\label{B4}
\end{eqnarray}

\begin{eqnarray}
\varphi(\mu _e,\mu _c,T)  = n^2_c \left(\frac{\partial n_e }{\partial \mu _e } \right)_{\mu _c,T}-
n_e n_c \left(\frac{\partial n_e }{\partial \mu _c } \right)_{\mu _e,T} -
n_e n_c \left(\frac{\partial n_c }{\partial \mu _e } \right)_{\mu _c,T} +
n_e^2 \left(\frac{\partial n_c }{\partial \mu _c } \right)_{\mu _e,T}.
\label{B5}
\end{eqnarray}

It should be noted that expressions (4) and (5) for the isothermal compressibility are fully consistent with the description of small-angle neutron and X-ray scattering [15-17].

From the limit relation (2) for the isothermal compressibility, taking into account (4) and (5), we come to the conclusion that the critical point in the two-component CS is possible in two cases [11]

{\it(A)} while simultaneously satisfying the limit relations
\begin{eqnarray}
\left(\frac{\partial n_e }{\partial \mu _e } \right)_{\mu _c,T}\to \infty \textrm{ è } \left(\frac{\partial n_c }{\partial \mu _c } \right)_{\mu _e,T} \to \infty ;
\label{B6}
\end{eqnarray}

{\it(B)} while satisfying the condition
\begin{eqnarray}
\varphi(\mu _e,\mu _c,T)\to  0.
\label{B7}
\end{eqnarray}

We now pay attention that conditions (6) correspond to the limit relation for determining the critical parameters of the L-V phase transition for the model single-component system with the short-range potential of the interparticle interaction (the so-called simple liquid) (see, e.g., [4]). The difference is that there is a single limit relation when considering the simple liquid, instead of two ones (6) for the CS which is initially a two-component system. Therefore, we can consider that conditions (6) correspond to the "ordinary" critical point of the L-V phase transition in the two-component CS.

Thus, we need to show that the condition (7) corresponds to the critical point associated with the M-NM phase transition. To consider the CS static conductivity $\sigma_{st}$, we will use the linear response theory within which the interparticle interaction can in principle be adequately considered (see, e.g., [9]).

According to a linear response theory, the quantity $\sigma_{st}$ is directly related to the long-wavelength limit of the longitudinal dielectric permittivity (DP) $\varepsilon^{l} (q,\omega)$ [18],
\begin{eqnarray}
\varepsilon(\omega ) = \lim_{q \to 0}\varepsilon^{l}(q,\omega )= 1+ \frac{4\pi  i}{\omega }\sigma (\omega ), \qquad \sigma _{st}=\lim_{\omega  \to 0}\sigma (\omega ).
\label{B8}
\end{eqnarray}
In this case, the longitudinal permittivity is given by
\begin{eqnarray}
\varepsilon^{l}(q,\omega )= \left\{1+ \frac{4\pi }{q^2} \chi ^R(q,\omega )     \right\}^{-1},
\label{B9}
\end{eqnarray}
where $\chi^R(q,\omega)$ is the retarded Green's function "charge-charge"
\begin{eqnarray}
\chi ^R(q,\omega )= \int_{0}^{\infty} dt\exp(i\omega t)f_{\chi }(q,t), \qquad f_{\chi }(q,t)= - \frac{i}{\hbar V} \langle [\rho_q(t),\rho_{-q}(0)]\rangle .
\label{B10}
\end{eqnarray}

Here $\rho_q(t)$ is the Fourier component of the charge density operator in the Heisenberg representation, angle brackets denote averaging with the grand canonical distribution. Relation (10) should be understood in the thermodynamic limit. From (8) - (10), using the quantum field theory methods [14], it immediately follows [19] that the dynamic conductivity is defined by the known Kubo formula [20].
\begin{figure}
\includegraphics[width=8cm]{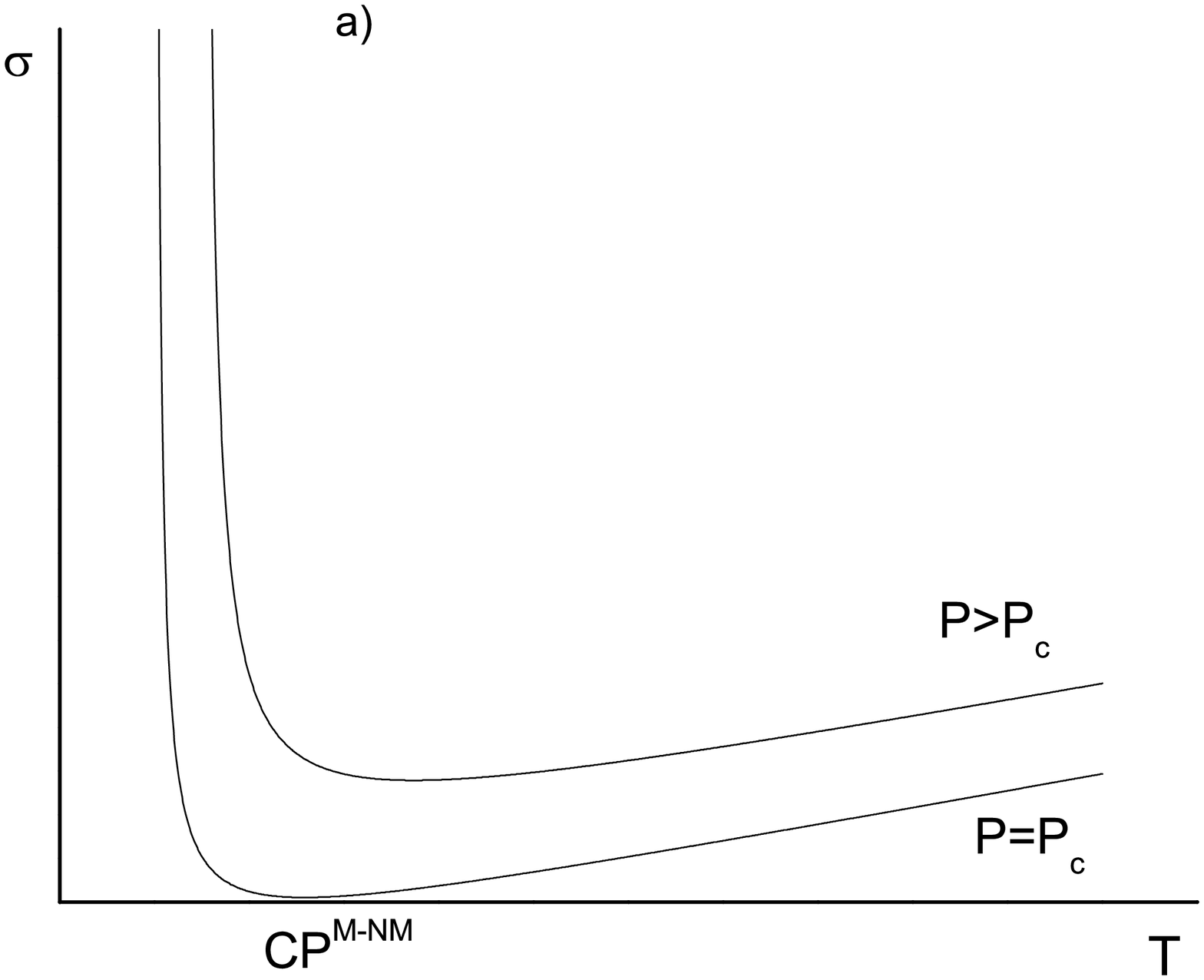}
\includegraphics[width=8cm]{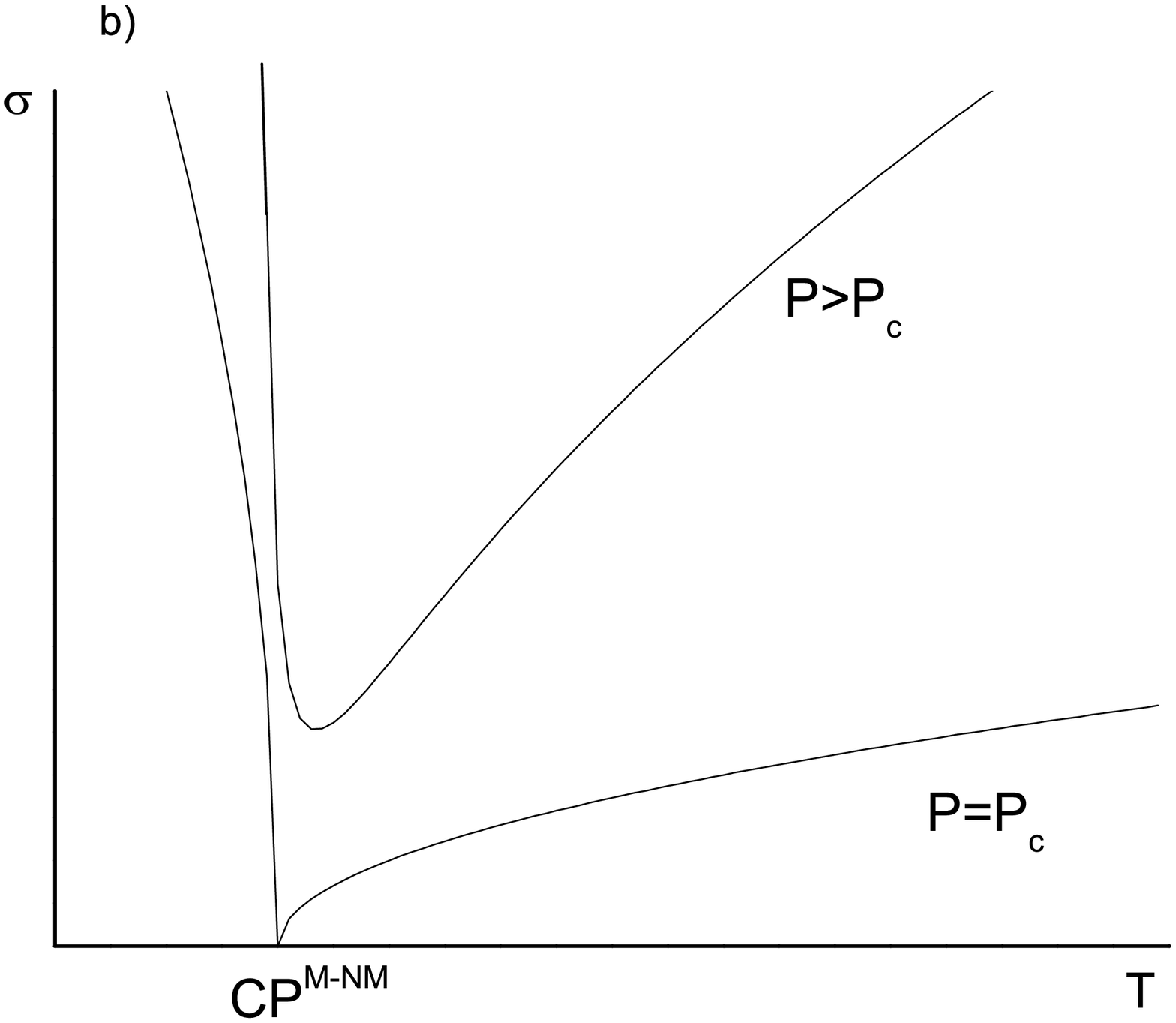}
\caption{The possible variants a) and b) of qualitative behavior of the static conductivity in the vicinity of the second critical point CP$^{M-NM}$, according to the present theory (see also [11]).}
\label{fig.2}
\end{figure}
Thus, the dynamic permittivity $\varepsilon(\omega)$ has a $\omega$-singularity in the static limit, which is related to the static conductivity,
\begin{eqnarray}
\varepsilon(\omega )|_{\omega \to 0}\to \frac{4\pi i}{\omega }\sigma _{st},
\label{B11}
\end{eqnarray}
At the same time, for the static permittivity $\varepsilon(q) = \lim_{\omega \to 0}\varepsilon(q,\omega) $, the limit relation [15] is satisfied
\begin{eqnarray}
\lim_{q \to 0}q^2 \{\varepsilon(q)-1 \}= \kappa^2, \qquad \kappa^2= 4\pi \sum_{a,b} z_az_b e^2  \left(\frac{\partial n_a }{\partial \mu _b } \right)_{\mu _c,T},
\label{B12}
\end{eqnarray}
where $\kappa$, being the quantity inverse to the screening length, characterizes the length of static external field penetration into the CS at arbitrary thermodynamic parameters in the normal unordered CS.
\begin{figure}[h]
\centering\includegraphics[width=10cm]{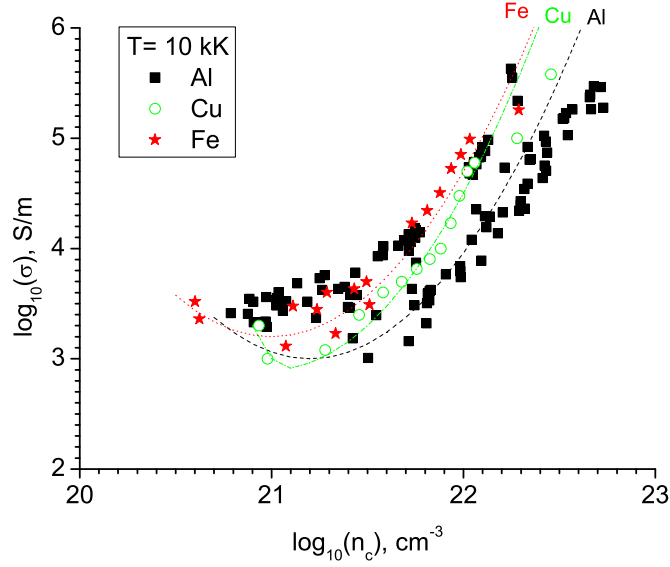}
\caption{{\protect\footnotesize {The experimental data [2], [5], [21] for the density dependence of the static conductivity for Al, Cu and Fe, which demonstrate minimum $\sigma_{st}$ and, according to the presented theory, placed in the vicinity of CP$^{M-NM}$.}}}\label{Fig.3}
\end{figure}
Thus, the longitudinal permittivity $\varepsilon^{l}(q, \omega)$ is characterized by two singularities: 1$/\omega $ in the static limit $\omega \to 0$ of the dynamic permittivity $\varepsilon(\omega)$ (see (11)) and the singularity 1$/q^2$ in the long-wavelength limit $q\to 0$ of the static permittivity $\varepsilon(q)$ (see (12)). As shown in [12], these singularities of the longitudinal permittivity are one-to-one related. In particular, this means that the disappearance of one of the singularities results in the disappearance of the other,
\begin{eqnarray}
\sigma _{st}\to 0  \leftrightarrow  \kappa\to 0.
\label{B13}
\end{eqnarray}
It follows from relation (13) that the screening length of the Coulomb interparticle interaction is larger than any characteristic size in the state of "true" dielectric with zero static conductivity, and the screening effect is reduced only to a change in the interaction amplitude [11,12]. As a result, it can be considered that charged particles are in the collective "localized" state with zero static conductivity, which is caused by the "divergence" at long distances between particles.
\begin{figure}[h]
\centering\includegraphics[width=10cm]{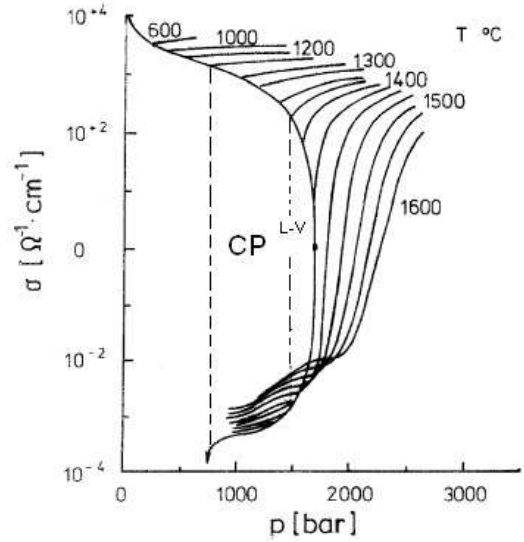}
\caption{{\protect\footnotesize {The experimental dependence of the static electrical conductivity $\sigma_{st}$ on pressure along the isotherms in mercury (see [5], [22],[23] and references therein). The intersection of the isotherms for conductivity confirms existence of the CP$^{M-NM}$. It is necessary to underline that in reality the continuous curves for conductivity on this figure are obtained from the experimental points (similar to the results on Fig.3) and the smooth character of these curves can be violated for some values of the thermodynamic parameters.}}}\label{fig.4}
\end{figure}
We now pay attention that the isothermal compressibility $K_T$ is a physical quantity; therefore, in relations (4) and (5), we can use the quasi-neutrality condition (1) for densities, rather than their derivatives. Thus, we obtain
\begin{eqnarray}
\varphi(\mu _e,\mu _c,T)= n^2_c \left\{
\left(\frac{\partial n_e }{\partial \mu _e } \right)_{\mu _c,T} -
z_c\left(\frac{\partial n_e }{\partial \mu _c } \right)_{\mu _e,T}-
z_c\left(\frac{\partial n_c }{\partial \mu _e } \right)_{\mu _c,T}+
z_c^2\left(\frac{\partial n_c }{\partial \mu _c } \right)_{\mu _e,T}
\right\}.
\label{B14}
\end{eqnarray}

Taking into account relations (12) and (14), it follows from (13) that
\begin{eqnarray}
\sigma _{st}\to 0  \leftrightarrow  \varphi (\mu_e, \mu_c, T) \to 0.
\label{B15}
\end{eqnarray}
We come to the conclusion that expanded metals in the second critical point associated with the M-NM phase transition are in the state of true dielectric (the state with zero conductivity) (see Fig. 2). According to the present theory (see also [11]) in the point CP$^{M-NM}$ the substance is in a specific state which we named the "true dielectric" (or "absolute dielectric"). In this state the static conductivity equals zero. Conductivity near the transition characterizes by temperature dependence $\sigma_{st}\sim (T-T_c^{M-NM})^\gamma$: in the case a) with $\gamma>1$; in the case b) $1>\gamma>0$ (there is singularity in the temperature derivative of $\sigma_{st}$). The similar pictures can be drawn for the dependence of $\sigma$ from other variables as, e.g., pressure, or density, since the thermodynamic process which leads to the CP$^{M-NM}$ can be chosen arbitrary.

The experimental data on the conductivity of various metals [5,21] confirm this conclusion (see Fig. 3 for Al, Cu and Fe  and Fig. 4 for mercury). The position of the local minimum of the conductivity characterizes the approximate value of the corresponding thermodynamic parameter of the second critical point in expanded metals. In mercury, according to the experimental data [5], we can suppose that $T_c^{M-NM}>T_c^{L-VP}$ and, therefore, the case b) on Fig. 1 is realized (with $P_c^{M-NM}>P_c^{L-VP}$).

Thus, based on the above consideration and available experimental data, we can argue that there are two critical points in expanded metals. One of the critical points corresponds to the ordinary L-V phase transition; the second critical point corresponds to the M-NM phase transition. It is shown that if the second critical point exists, the CS at this point is in the state of true dielectric with zero static conductivity. The results of this paper give the general and explicit explanation of the opportunity for the appearance of the second critical point and show the way of it possible calculations. The developed consideration will be further applied to other elements, including hydrogen and the noble gases.

\section*{Acknowledgment}

This study was supported by the Netherlands Organization for Scientific Research (NWO) and by joint grant of the Russian Foundation for Basic Research and National Academy of Science Ukraine, projects no. 12-08-00600-à and no. 12-02-90433-Ukr-a. The authors are thankful to E.M. Apfelbaum, W. Ebeling, G.J.F. van Heijst and A.D. Rakhel for useful discussions.

\end{document}